# Measurement of the probe Stark shift of the $^{87}$Sr optical lattice clock using the frequency modulation spectroscopy

**QINFANG XU,**[1,2] **XIAOTONG LU,**[1,2] **JINGJING XIA,**[1] **YEBING WANG,**[1] **HONG CHANG,**[1,2]*

[1]*CAS Key Laboratory of Time and Frequency Primary Standards, National Time Service Center, Xi'an 710600, China*
[2] *School of Astronomy and Space Science, University of Chinese Academy of Sciences, Beijing 100049, China*
*Corresponding author: changhong@ntsc.ac.cn*

**With the uncertainty of the optical clocks improving to the order of 10$^{-18}$, the probe light used to detect the clock transition has demonstrated nonnegligible Stark shift, provoking to precisely evaluate this shift. Here, we demonstrate a frequency modulation technique to realize a large measurement lever arm of the probe Stark shift with no cost of the measurement accuracy of the interleaved stabilization method. This frequency-modulated spectrum is theoretical described and experimental verified. The probe Stark shift coefficient of the $^{87}$Sr optical lattice clock is experimentally determined as -(45.97±3.51) Hz/(W/cm$^2$) using this frequency modulation spectroscopy.**

With two decades of development, optical clocks have demonstrated 10$^{-18}$ performance [1-5], and especially, optical lattice clocks show the 10$^{-19}$ stability benefitting of ultra-low quantum projection noise (QPN) [2,5]. These high-performing optical clocks can not only be used to redefine the second in the International System of Units (SI) [6,7], but also open new physics, such as detection of the dark matter [8], measurement of the time variation in fundamental constants [9,10], test of the Lorentz symmetry [11].

With the 10$^{-18}$ uncertainty of optical clocks, the probe Stark shift (PSS) is usually non-negligible [3-4]. When the Ramsey detection is used, although the two short π/2-pulses cause large PSS due to large probe light intensity, fortunately, the auto-balance Ramsey spectrum [12-14] and hyper-Ramsey spectroscopy [15] can reduce the probe light shift, and the residual PSS can be controlled below the 10$^{-18}$ level in terms of optical clocks. However, these artful Ramsey techniques should precisely tailor the phases, frequencies, and durations of the probe light pulse, or increase the clock feedback cycle decreasing the stability. In contrast, as the Rabi detection is used [5], this shift needs to be carefully evaluated such as using the interleaved stabilization method. At present, the Rabi detection is most adoptive method in optical clocks due to the simple detection process and lower PSS than Ramsey detection [2, 4, 12]. To precisely determine the PSS of the Rabi detection typically needs to make the intensity difference between two interleaved loops be as large as possible to create large measurement lever arm. However, to maintain a π-pulse of both interleaved loops, the larger probe light intensity corresponds to the shorter pulse duration, leading to a broader spectrum, which eventually limits the accessible measurement lever arm as the measurement precision decreases with the broadening spectrum.

In this paper, we demonstrate a measurement of PSS of $^{87}$Sr optical lattice clock by frequency modulation spectroscopy in detail. With this method, the coupling strength of the probe light to the clock transition can be well controlled by adjusting the modulation index without changing the total probe light intensity and thus the contradiction between the large measurement lever arm and broad spectrum is resolved.

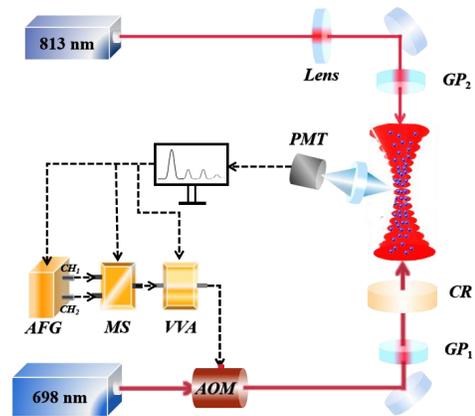

Fig. 1. Experimental setup of clock transition detection and frequency modulation of the probe light. PMT, photomultiplier; AOM, acousto-optic modulation; GP$_{1-2}$, Glan-Taylor polarizers; CR, concave mirror; AFG, arbitrary function generator; MS, microwave switch; VVA, voltage-

controlled attenuator. The MS and VVA are controlled by a computer that also processes the data collected by PMT, computes the frequency correction, and corrects the output frequency of the AFG to keep the probe light resonating with the atoms. The beam waist of the 698 nm probe laser is 705(10) μm, and overlaps with the waist of the 813 nm lattice laser (100 μm).

Figure 1 shows the schematic setup of the clock transition detection and probe light frequency modulation. After two-stage laser cooling, cold atoms are loaded into a horizontal one-dimensional optical lattice. The lattice laser wavelength of 813.42 nm is stabilized to an ultra-low-expansion (ULE) cavity by the Pound-Drever-Hall (PDH) technique. The lattice is formed by overlapping the linear-polarized incident beam and its retroreflected beam without any polarization rotation. By using the 689.44 nm transition of $\left|^1S_0, F=9/2\right\rangle \to \left|^3P_1, F=9/2\right\rangle$, about $10^4$ atoms are optically pumped into one of the $\left|^1S_0, m_F=\pm 9/2\right\rangle$ stretched states.

With state preparation complete, the clock transition is performed on the 698 nm $^1S_0 \to {}^3P_0$ transition using the probe laser of which the fractional stability is $1\times 10^{-15}$ @1 s after PDH stabilizing to a 10 cm ULE cavity with a finesse of 40,0 000 [16]. A bias magnetic field, which defines the quantization axis along the direction of the gravity, splits the $m_F=\pm 9/2$ states by about 400 Hz corresponding to a magnetic field magnitude of 410 mG [17], and the clock transition is interrogated with a π-pulse. The polarizations of both the lattice and probe lasers are collinear with the quantization axis using the Glan-Taylor polarizer. The excitation fraction of the clock transition is detected by the electron-shelving method [18], where the 461 nm transition is used to detect the population in $^1S_0$ state and the fluorescence is collected by a photomultiplier (PMT).

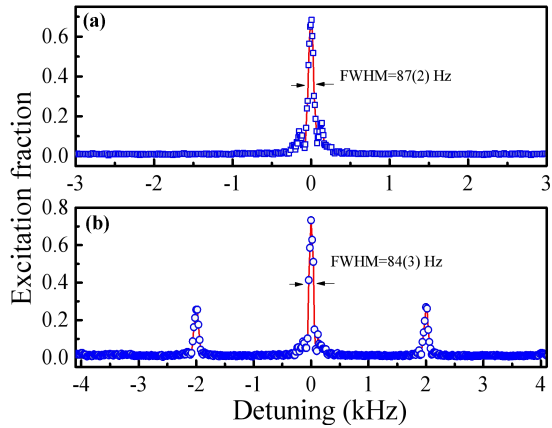

Fig. 2. Clock transition spectra comparison. (a) The spectrum obtained by 10 ms π-pulse without frequency modulation. (b) The frequency-modulated spectrum measured with the same clock laser duration and intensity with (a). The modulation amplitude and frequency are 0.8 kHz and 2 kHz, respectively.

The interleaved stabilization technique is used to measure the frequency difference of the two interleaved clock loops [19-21]. One loop operates under low probe light power so that the 100 ms π-pulse is realized. The other loop runs with much larger probe light power and the frequency modulation technique is used to make sure a 100 ms π-pulse. The lock-in data of each interleaved loops are the average frequency between the $m_F=\pm 9/2 \to m_F=\pm 9/2$ transitions, and the lock-in data frequency difference between the two loops represents the PSS with corresponding probe intensity difference. A two-channels arbitrary function generator (AFG), stabilized to a hydrogen clock, is used to generate the frequency shift signals of the two interleaved loops, where one of the channels output standard sinusoidal signal, the other output a frequency-modulated sinusoidal signal. A microwave switch and voltage-controlled attenuator are used to, respectively, switch the two signals and control the power by changing the diffraction efficiency of the acousto-optic modulation (AOM).

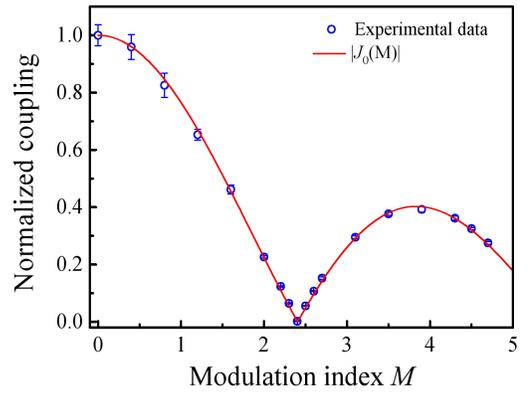

Fig. 3. Comparison of the experimental measurements of the normalized coupling strength of the 0th order sideband and the absolute value of the first-class zero-order Bessel function at different modulation index $M$. The error bars represent the 1σ standard of three independent measurements.

With modulating, the probe light frequency becomes $\omega_p(t)=\omega_p+\omega_a \cos(\omega_s t)$ after passing the AOM, where $\omega_p$ is probe light frequency without frequency modulation. After the rotating wave approximation, the Hamiltonian of the two energy levels atomic gas can be described by [22]

$$\hat{H}_{n_z,n_r}(t) = \frac{\hbar}{2}(\delta + \omega_a \cos(\omega_s t))\hat{\sigma}^{(3)}_{n_z,n_r} + \frac{\hbar g_{n_z,n_r}}{2}\hat{\sigma}^{(1)}_{n_z,n_r}. \quad (2)$$

Where $\delta = \omega_c - \omega_p$ is the detuning, $\omega_c$ is the clock transition frequency of $^{87}$Sr; $n_z$ and $n_r$ are the longitudinal and transverse motional quantum numbers of the lattice-trapped atoms, respectively; $g_{n_z,n_r} = g_0 e^{-(\eta_z^2+\eta_r^2)/2} L_{n_z}(\eta_z^2) L_{n_r}(\eta_r^2)$ is the modified coupling strength of probe light to the clock transition in the external states $n_z$, $n_r$, where $g_0$ is the bare states coupling strength and can be extracted from the Rabi oscillation, $\eta_z=0.269$ and $\eta_r=0.043$ are the longitudinal and transverse Lamb-Dicke parameters, respectively [23], and the $L_n(\cdot)$ is 0th order generalized Laguerre polynomial; $\hat{\sigma}^{(1)}_{n_z,n_r}$ and $\hat{\sigma}^{(3)}_{n_z,n_r}$ are Pauli matrices. By the unitary operator $\hat{U}(t) = e^{-\frac{i}{2}(\delta t + \frac{\omega_a}{\omega_s}\sin\omega_s t)\hat{\sigma}_z}$, the interaction picture Hamiltonian can be written by [24]

$$\hat{H}'_{n_z,n_r}(t) = \frac{\hbar g_{n_z,n_r}}{2}[\sum_{n=-\infty}^{+\infty} J_n(M)e^{i(\delta+n\omega_s)t}\hat{\sigma}_+ + h.c.]. \qquad (2)$$

where $J_n$ is the Bessel function of the first kind, $M=\omega_a/\omega_s$ is the modulation index. Eq. (2) indicates that the frequency modulation generated a series of sidebands, of which the effective Rabi coupling strength for the $n$th sideband is $g_{n_z,n_r}J_n(M)$. As the experiment is operated in the parameter region where the resolved sideband approximation holds $g_{n_z,n_r}|J_n(M)| \ll \omega_s$ (the largest $g_{n_z,n_r}$ is less than 300×2π Hz, however the $\omega_s$ is 2000×2π Hz in this experiment), all the sidebands can be treated resolvable, as shown in Fig. 2. So, when $\delta \approx -n\omega_s$, we can neglect those non-resonant terms proportional to $e^{i(\delta+m\omega_s)t}$ with m≠n. Thus, we can independently deal with sidebands without considering the effects of other sidebands. In this experiment, we used the carrier ($n$=0) to measure the PSS by the interleaved stabilization method. The total PSS will keep unchanged as the differential polarizability between the $^1S_0$ and $^3P_0$ is linear and the sidebands are symmetrical about the carrier. We experimentally measured the effective coupling strength $g_{n_z,n_r}J_0(M)$ of the carrier as shown in Fig. 3. Well agreement between the experimental measurement and the theory shows that the frequency modulation index can be precisely controlled and the theoretical model can well describe our system.

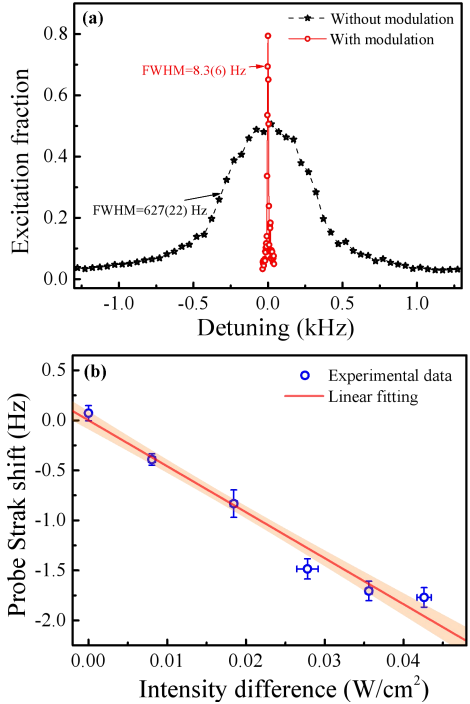

linewidth of 8.9 Hz, and about 0.5 excitation is the result of the dephasing effect of the probe laser and the atoms. The red circles represent the FWHM of 8.3(6) Hz, also using 100 ms pulse with the same probe light power with the black stars as the modulation frequency and modulation amplitude are 2 kHz and 4.7182 kHz, respectively. (b) The measured relationship between the PSS and probe light intensity difference between the two interleaved loops. The red solid line indicates the linear fitting, and the orange shaded region marks the 1σ uncertainty of the fitting line.

To precisely determine the PSS, large intensity difference of the probe light between the two interleaved loops is needed. Meanwhile, the full width at half maximum (FWHM) of the spectrum should keep unchanged, indicating that the pulse duration of the probe light of the two interleaved loop is constant. Benefiting from that the frequency-modulated spectrum can discretionarily adjust the effective coupling strength of sidebands as the value of the $|J_n(M)|$ ranges from 0 to 1, indicating that for any probe light intensity can be arbitrarily adjusted by the value of $|J_n(M)|$, and thus, the pulse duration can always keep constant. In this experiment, the maximum probe light intensity is 0.043(1) W/cm$^2$ and the corresponding spectrum, with the full width at half maximum (FWHM) of 627(22) Hz broadened by the high probe light intensity, is shown by the black stars in Fig. 4(a) when the 100 ms pulse is used (the corresponding Fourier limited FWHM is about 8 Hz). Compared with the several-hundred linewidth, the frequency-modulated spectrum of the carrier (the 0th sideband) demonstrates a linewidth of 8.3(6) Hz with the same intensity and pulse duration.

Figure 4(b) shows the experimental result of the PSS in an $^{87}$Sr optical lattice clock using the frequency-modulated spectrum. Each data is obtained by lock-in measurement using the interleaved stabilization technique. The $y$-error bars represent the statistical uncertainty given by the last point of the Allan deviation of the lock-in data. The $x$-error bars are given by the 1σ standard of the probe light intensity measurement during the corresponding measurement process in Fig. 4(b). By linear fitting (using the $y$-error bars as the weight), the slope of the experimental data is -(45.97±3.51) Hz/(W/cm$^2$), where the uncertainty indicates 1σ standard of the fitting slope.

Fig. 4. The measurement of the probe light shift of $^{87}$Sr optical lattice clock. (a) The comparison of the spectrum with/without frequency modulation. The black squares show a FWHM of 627(22) Hz when a 100 ms probe pulse is used (the intensity is 0.043 W/cm$^2$). The power broadening causes that the spectrum linewidth is larger than the corresponding Fourier transform limiting

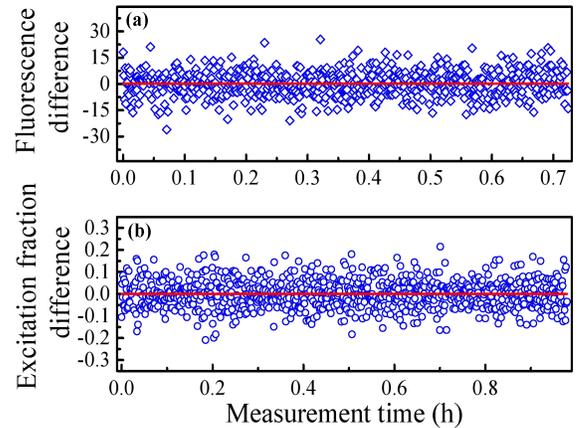

Fig. 5. The monitor of system parameters difference between two interleaved stabilization loops. (a) The fluorescence difference detected by the PMT, where the fluorescence is proportional to the total atomic number. The red solid line indicates the average value of 0.2(28). (b) The excitation fraction difference with a mean of -0.008(2) shown by the red solid line.

The density shift may change due to the atomic number drift and the difference of the average excitation fraction between the two loops [25-27] due to the imperfect parameter setting of the *M*. We monitor the fluorescence and the excitation fraction difference between the two loops as typically shown in Fig. 5(a) and (b), respectively. According to our previous measurement results in reference [28], for each data shown in Fig. 4(b), we correct the offset due to the density difference between the two interleaved loops. As the PSS has the tensor part and is relative to the angle between the bias magnetic field and the probe light polarization, thus, the magnitude of the magnetic field, extracted from the frequency gap between the $m_F$=+9/2 and $m_F$=-9/2 [11], is monitored. There is no observable magnetic field drift in this experiment, indicating that the error caused by the magnetic field drift can be neglected.

In conclusion, we describe the coupling strength control technique by frequency-modulated spectroscopy and demonstrate the measurement of the probe Stark shift by this method. By measuring the relationship between the probe Stark shift and relative intensity, the probe Stark shift coefficient of $^{87}$Sr optical lattice clock is determined as -(45.97±3.51) Hz/(W/cm$^2$), indicating that the probe Stark shift is -7.2(6)×10$^{-19}$ for the regular operation condition of our clock with a probe light intensity of 6.6×10$^{-6}$ W/cm$^2$ (with a staggering measurement lever arm of 6515). The measured probe Stark shift coefficient of $^{87}$Sr is quite different from $^{88}$Sr, which has been determined as -(13±2) Hz/(W/cm$^2$) [29], indicating that it is inaccurate to evaluate the systematic error of the probe Stark shift of the $^{87}$Sr optical lattice clock using the probe Stark shift coefficient of the $^{88}$Sr [30,31]. The technique shown in this paper can be used in any clock to much improve the measure precision of the probe Stark shift for meeting the requirement of continuous improved clocks.

**Funding.** This research was supported by the National Natural Science Foundation of China (No. 61775220), the Key Research Project of Frontier Science of the Chinese Academy of Sciences (Grant No. QYZDB-SSW-JSC004), and the Strategic Priority Research Program of the Chinese Academy of Sciences (Grant No. XDB35010202).

**Disclosures.** The authors declare no conflicts of interest.

**Data availability.** Data underlying the results presented in this paper are not publicly available at this time but may be obtained from the authors upon reasonable request.

## References

1. S. M. Brewer, J. S. Chen, A. M. Hankin, E. R. Clements, C. W. Chou, and D. J. Wineland, Phys. Rev. Lett. 123, 033201 (2019)
2. W. F. McGrew, X. Zhang, r. J. Fasano, S. A. Schäffer, K. Beloy, D. Nicolodi, r. C. Brown, N. Hinkley, G. Milani, M. Schioppo, t. H. Yoon, and A. D. Ludlow, Nature. 564, 87 (2018).
3. B. J. Bloom, T. L. Nicholson, J. R. Williams, S. L. Campbell, M. Bishof, X. Zhang, W. Zhang, S. L. Bromley, and J. Ye, Nature. 506, 71 (2014).
4. T. L. Nicholson1, S. L. Campbell1, R. B. Hutson2, G. E. Marti1, B. J. Bloom, R. L. McNally, W. Zhang, M. D. Barrett, M. S. Safronova, G. F. Strouse, W. L. Tew, and J. Ye, Nat. Commun. 6, 6896 (2015).
5. T. Bothwell, D. Kedar, E. Oelker, J. M. Robinson, S. L. Bromley, W. L. Tew, J. Ye, and C. J. Kennedy, Metrologia. 56, 065004 (2019).
6. F. Riehle, C. R. Phys. 16, 506 (2015).
7. S. Bize, C. R. Phys. 20 153–68 (2019).
8. A. Derevianko, and M. Pospelov, Nat. Phys. 10, 933 (2014).
9. N. Huntemann, B. Lipphardt, C. Tamm, V. Gerginov, S. Weyers, and E. Peik, Phys. Rev. Lett. 113, 210802 (2014).
10. R. M. Godun, P. Nisbet-Jones, J. Jones, S. King, L. Johnson, H. Margolis, K. Szymaniec, S. Lea, K. Bongs, and P. Gill, Phys. Rev. Lett. 113, 210801 (2014).
11. C. Sanner, N. Huntemann, R. Lange, c. tamm, E. Peik, M. S. Safronova, and S. G. Porsev, Nature. 567, 204 (2019).
12. C. Sanner, N. Huntemann, R. Lange, C. Tamm, and E. Peik, Phys. Rev. Lett. 120, 053602 (2018).
13. V. I. Yudin, A. V. Taichenachev, M. Yu. Basalaev,1 T. Zanon-Willette, J. W. Pollock, M. Shuker, E. A. Donley, and J. Kitching, Phys. Rev. Applied. 9, 054034 (2018).
14. M. Shuker, J. W. Pollock, R. Boudot, V. I. Yudin, A. V. Taichenachev, J. Kitching, and E. A. Donley, Phys. Rev. Lett. 122, 113601 (2019).
15. K. Beloy, Phys. Rev. A. 97, 031406(R) (2018).
16. X. T. Lu, C. H. Zhou, T. Li, Y. B. Wang, and H. Chang, Appl. Phys. Lett. 117, 231101 (2020).
17. M. M. Boyd, T. Zelevinsky, A. D. Ludlow, S. Blatt, T. Zanon-Willette, S. M. Foreman, and J. Ye, Phys. Rev. A. 76, 022510 (2007).
18. W. Nagourney, I. Sandberg, and H. Dehmelt, Phys. Rev. Lett. 56, 2797 (1986).
19. A. Al-Masoudi, S. Dörscher, S. Häfner, U. Sterr, and Ch. Lisdat, Phys. Rev. A. 92, 063814 (2015).
20. Y. Li, Y. G. Lin, Q. Wang, T. Yang, Z. Sun, E. J. Zang, and Z. J. Fang, Chin. Opt. Lett. 16, 051402 (2018).
21. T. L. Nicholson, M. J. Martin, J. R. Williams, B. J. Bloom, M. Bishof, M. D. Swallows, S. L. Campbell, and J. Ye, Phys. Rev. Lett. 109, 230801 (2012).
22. L. H. Huang, P. Peng, D. H. Li, Z. M. Meng, L. C. Chen, C. L. Qu, P. J. Wang, C. W. Zhang, and J. Zhang, Phys. Rev. A. 98, 013615 (2018).
23. S. Blatt, J. W. Thomsen, G. K. Campbell, A. D. Ludlow, M. D. Swallows, M. J. Martin, M. M. Boyd, and J. Ye, Phys. Rev. A. 80, 052703 (2009).
24. M. P. Silveri, J. A. Tuorila, E. V. Thuneberg, and G. S. Paraoanu, Rep. Prog. Phys. 80, 056002 (2017).
25. A. M. Rey, A. V. Gorshkov, and C. Rubbo, Phys. Rev. Lett. 103, 260402 (2009).
26. N. D. Lemke, J. von Stecher, J. A. Sherman, A. M. Rey, C. W. Oates, and A. D. Ludlow, Phys. Rev. Lett. 107, 103902 (2011).
27. S. Lee, C. Y. Park, W. K. Lee, and D. H. Yu, New J. Phys. 18, 033030 (2016).
28. C. H. Zhou, X. T. Lu, B. Q. Lu, Y. B. Wang, and H. Chang, Appl. Sci. 11, 1206 (2021).
29. X. Baillard, M. Fouché, R. L. Targat, P. G. Westergaard, A. Lecallier, Y. Le Coq, G. D. Rovera, S. Bize, and P. Lemonde, Opt. Lett. 32, 1812 (2007)
30. St. Falke, H. Schnatz, J. S. R. Vellore Winfred, Th. Middelmann, St. Vogt, S. Weyers, B. Lipphardt, G. Grosche, F. Riehle, U. Sterr, and Ch Lisdat, Metrologia. 48, 399 (2011).
31. H. Hachisu, and T. Ido, Jpn. J. Appl. Phys. 54, 112401 (2015).

## References

1. S. M. Brewer, J. S. Chen, A. M. Hankin, E. R. Clements, C. W. Chou, and D. J. Wineland, "$^{27}Al^+$ Quantum-Logic Clock with a Systematic Uncertainty below $10^{-18}$," Phys. Rev. Lett. 123, 033201 (2019)
2. W. F. McGrew, X. Zhang, r. J. Fasano, S. A. Schäffer, K. Beloy, D. Nicolodi, r. C. Brown, N. Hinkley, G. Milani, M. Schioppo, t. H. Yoon, and A. D. Ludlow, "Atomic clock performance enabling geodesy below the centimetre level," Nature. 564, 87 (2018).
3. B. J. Bloom, T. L. Nicholson, J. R. Williams, S. L. Campbell, M. Bishof, X. Zhang, W. Zhang, S. L. Bromley, and J. Ye, "An optical lattice clock with accuracy and stability at the $10^{-18}$ level," Nature. 506, 71 (2014).
4. T. L. Nicholson1, S. L. Campbell1, R. B. Hutson2, G. E. Marti1, B. J. Bloom, R. L. McNally, W. Zhang, M. D. Barrett, M. S. Safronova, G. F. Strouse, W. L. Tew, and J. Ye, "Systematic evaluation of an atomic clock at $2\times10^{-18}$ total uncertainty," Nat. Commun. 6, 6896 (2015).
5. T. Bothwell, D. Kedar, E. Oelker, J. M. Robinson, S. L. Bromley, W. L. Tew, J. Ye, and C. J. Kennedy, "JILA SrI optical lattice clock with uncertainty of $10^{-18}$," Metrologia. 56, 065004 (2019).
6. F. Riehle, "Towards a redefinition of the second based on optical atomic clocks," C. R. Phys. 16, 506 (2015).
7. S. Bize, C. R. Phys. "The unit of time: Present and future directions," 20 153–68 (2019).
8. A. Derevianko, and M. Pospelov, "Hunting for topological dark matter with atomic clocks," Nat. Phys. 10, 933 (2014).
9. N. Huntemann, B. Lipphardt, C. Tamm, V. Gerginov, S. Weyers, and E. Peik, "Improved Limit on a Temporal Variation of $m_p/m_e$ from Comparisons of $Yb^+$ and Cs Atomic Clocks," Phys. Rev. Lett. 113, 210802 (2014).
10. R. M. Godun, P. Nisbet-Jones, J. Jones, S. King, L. Johnson, H. Margolis, K. Szymaniec, S. Lea, K. Bongs, and P. Gill, "Frequency Ratio of Two Optical Clock Transitions in $^{171}Yb^+$ and Constraints on the Time Variation of Fundamental Constants," Phys. Rev. Lett. 113, 210801 (2014).
11. C. Sanner, N. Huntemann, R. Lange, C. tamm, E. Peik, M. S. Safronova, and S. G. Porsev, "Optical clock comparison for Lorentz symmetry testing," Nature. 567, 204 (2019).
12. C. Sanner, N. Huntemann, R. Lange, C. Tamm, and E. Peik, "Autobalanced Ramsey Spectroscopy," Phys. Rev. Lett. 120, 053602 (2018).
13. V. I. Yudin, A. V. Taichenachev, M. Yu. Basalaev,1 T. Zanon-Willette, J. W. Pollock, M. Shuker, E. A. Donley, and J. Kitching, "Generalized Autobalanced Ramsey Spectroscopy of Clock Transitions," Phys. Rev. Applied. 9, 054034 (2018).
14. M. Shuker, J. W. Pollock, R. Boudot, V. I. Yudin, A. V. Taichenachev, J. Kitching, and E. A. Donley, "Ramsey Spectroscopy with Displaced Frequency Jumps," Phys. Rev. Lett. 122, 113601 (2019).
15. K. Beloy, "Hyper-Ramsey spectroscopy with probe-laser-intensity fluctuations," Phys. Rev. A. 97, 031406(R) (2018).
16. X. T. Lu, C. H. Zhou, T. Li, Y. B. Wang, and H. Chang, "Synchronous frequency comparison beyond the Dick limit based on dual-excitation spectrum in an optical lattice clock," Appl. Phys. Lett. 117, 231101 (2020).
17. M. M. Boyd, T. Zelevinsky, A. D. Ludlow, S. Blatt, T. Zanon-Willette, S. M. Foreman, and J. Ye, "Nuclear spin effects in optical lattice clocks," Phys. Rev. A. 76, 022510 (2007).
18. W. Nagourney, I. Sandberg, and H. Dehmelt, "Shelved Optical Electron Amplifier: Observation of Quantum Jumps," Phys. Rev. Lett. 56, 2797 (1986).
19. A. Al-Masoudi, S. Dörscher, S. Häfner, U. Sterr, and Ch. Lisdat, "Noise and instability of an optical lattice clock," Phys. Rev. A. 92, 063814 (2015).
20. Y. Li, Y. G. Lin, Q. Wang, T. Yang, Z. Sun, E. J. Zang, and Z. J. Fang, "An improved strontium lattice clock with 10–16 level laser frequency stabilization," Chin. Opt. Lett. 16, 051402 (2018).
21. T. L. Nicholson, M. J. Martin, J. R. Williams, B. J. Bloom, M. Bishof, M. D. Swallows, S. L. Campbell, and J. Ye, "Comparison of Two Independent Sr Optical Clocks with $1\times10^{-17}$ Stability at $10^3$ s," Phys. Rev. Lett. 109, 230801 (2012).
22. L. H. Huang, P. Peng, D. H. Li, Z. M. Meng, L. C. Chen, C. L. Qu, P. J. Wang, C. W. Zhang, and J. Zhang, "Observation of Floquet bands in driven spin-orbit-coupled Fermi gases," Phys. Rev. A. 98, 013615 (2018).
23. S. Blatt, J. W. Thomsen, G. K. Campbell, A. D. Ludlow, M. D. Swallows, M. J. Martin, M. M. Boyd, and J. Ye, "Rabi spectroscopy and excitation inhomogeneity in a one-dimensional optical lattice clock," Phys. Rev. A. 80, 052703 (2009).
24. M. P. Silveri, J. A. Tuorila, E. V. Thuneberg, and G. S. Paraoanu, "Quantum systems under frequency modulation," Rep. Prog. Phys. 80, 056002 (2017).
25. A. M. Rey, A. V. Gorshkov, and C. Rubbo, "Many-Body Treatment of the Collisional Frequency Shift in Fermionic Atoms," Phys. Rev. Lett. 103, 260402 (2009).
26. N. D. Lemke, J. von Stecher, J. A. Sherman, A. M. Rey, C. W. Oates, and A. D. Ludlow, "*p*-Wave Cold Collisions in an Optical Lattice Clock," Phys. Rev. Lett. 107, 103902 (2011).
27. S. Lee, C. Y. Park, W. K. Lee, and D. H. Yu, "Cancellation of collisional frequency shifts in optical lattice clocks with Rabi spectroscopy," New J. Phys. 18, 033030 (2016).
28. C. H. Zhou, X. T. Lu, B. Q. Lu, Y. B. Wang, and H. Chang, "Demonstration of the Systematic Evaluation of an Optical Lattice Clock Using the Drift-Insensitive Self-Comparison Method," Appl. Sci. 11, 1206 (2021).
29. X. Baillard, M. Fouché, R. L. Targat, P. G. Westergaard, A. Lecallier, Y. Le Coq, G. D. Rovera, S. Bize, and P. Lemonde, "Accuracy evaluation of an optical lattice clock with bosonic atoms," Opt. Lett. 32, 1812 (2007)
30. St. Falke, H. Schnatz, J. S. R. Vellore Winfred, Th. Middelmann, St. Vogt, S. Weyers, B. Lipphardt, G. Grosche, F. Riehle, U. Sterr, and Ch Lisdat, "The $^{87}Sr$ optical frequency standard at PTB," Metrologia. 48, 399 (2011).
31. H. Hachisu, and T. Ido, "Intermittent optical frequency measurements to reduce the dead time uncertainty of frequency link," Jpn. J. Appl. Phys. 54, 112401 (2015).